\documentclass[aip,jcp,preprint,floatfix]{revtex4-1}

\usepackage{graphicx}
\usepackage{color}
\usepackage{dcolumn}
\usepackage{amsmath,amsfonts} %
\usepackage{mathtools}
\usepackage{braket} %
\usepackage[acronym]{glossaries} %
\usepackage[breaklinks, %
colorlinks, %
linkcolor=blue, %
citecolor=blue, %
urlcolor=blue]{hyperref}
\usepackage{comment}

\DeclareMathOperator{\tr}{Tr} %
\newcommand{\eq}[1]{Eq.~(\ref{#1})} %
\newcommand{\bea}{\begin{eqnarray}}
\newcommand{\eea}{\end{eqnarray}}

\newcommand{\op}[1]{\ensuremath{\hat{#1}}}
\renewcommand{\ket}[1]{\ensuremath{\left|#1\right\rangle}}
\renewcommand{\bra}[1]{\ensuremath{\left\langle #1\right|}}

\newcommand{\braOket}[3]{\ensuremath{\left\langle #1\middle|#2\middle|#3\right\rangle}}
\renewcommand{\Re}{\operatorname{Re}}

\newcommand{\e}[1]{\ensuremath{\, \mathrm{e}^{#1}}}
\newcommand\norm[1]{\left\lVert#1\right\rVert}

\newcommand{\PL}[0]{\mathit{\Pi}}

\newacronym{NESS}{NESS}{non-equilibrium steady state}

\begin{document}

\title{Quantum Kinetic Rates within the Nonequilibrium Steady State}



\author{Lo{\"i}c Joubert-Doriol}
\affiliation{Univ Gustave Eiffel, Univ Paris Est Creteil, CNRS, UMR 8208, MSME, F-77454 Marne-la-Vallée, France}
\author{Kenneth A. Jung}
\affiliation{Chemical Physics Theory Group, Department of Chemistry, and Center for Quantum Information and Quantum Control, University of Toronto,Toronto,Ontario M5S 3H6, Canada}
\author{Artur F. Izmaylov}
\affiliation{Chemical Physics Theory Group, Department of Chemistry, and Center for Quantum Information and Quantum Control, University of Toronto,Toronto,Ontario M5S 3H6, Canada}
\author{Paul Brumer}
\affiliation{Chemical Physics Theory Group, Department of Chemistry, and Center for Quantum Information and Quantum Control, University of Toronto,Toronto,Ontario M5S 3H6, Canada}

\date{\today}
\begin{abstract}
The  nonequilibrium  steady state  (NESS) of a quantum network
is central to a host  of  physical  and  biological  scenarios.
Examples include natural processes such as vision and photosynthesis, as well as technical
devices such as photocells, both activated by incoherent light (e.g. sunlight)
and leading to quantum transport.  Here, a completely
general  approach  to  defining components of a quantum network in the NESS, and obtaining  rates  of  processes \textit{between}
these components is  provided. Quantum  effects are explicitly included throughout, both in
(a) defining network components via projection operators, and (b) in
determining the role of coherences in rate processes. As
examples, the  methodology  is  applied to model cases, two versions of the
V-level system, and to the spin-boson model, wherein the role of the environment and
of internal system properties in determining the rates is examined.  In addition, the role of Markovian
vs. non-Markovian contributions is quantified, exposing conditions under which
NESS rates can be obtained by perturbing the nonequilibrium steady state.
\end{abstract}
\maketitle

\section{Introduction}

Quantum  networks, i.e., collections of interacting states that
are  responsible  for  the  transport of energy and matter, are
ubiquitous      in      technology      and      in      nature			
\cite{Roden2016,Wu2021,Ghasemi2020,Patil2021}.
Of particular importance are networks in the nonequilibrium steady state (NESS) driven for example by incident incoherent light (e.g.
  solar radiation). Examples include biophysically significant chemistry\cite{Ottolenghi1982,Wand2013} such as photosynthesis,
  or vision in molecular biology\cite{Polli2010-dp,Tscherbul2014-fq} or photo and solar cells in device  physics\cite{Jin2010,Piontkowski2018}.
In such processes, incoherent light is constantly being applied, the photoproduct continuously removed, and the 
initial state regenerated. The result is an NESS displaying (time independent) transport.
Quantum  {NESS}s are ubiquitous, but in need of further analysis and development.

The operation of the network, e.g., the rate at which the  total network produces product,
depends on the rates of transfer between components 
within the quantum network.
Hence obtaining NESS rates between constituents that form the quantum network is a general, significant,
challenge  and of  particular  relevance  to  the natural operating
conditions                                                   of
thermal~\cite{Segal2003,Li2012,Xu22016,Kilgour2018},
biological~\cite{Manzano2013,LenMontiel2014,ZerahHarush2018,Tscherbul2018,Jankovi2020,Yang2020,Jung2020}  and
electronic~\cite{Galperin2007,Navrotskaya2009,Subotnik2009,Zwolak2020}  systems
that exhibit transport of matter and/or energy.

Efforts have been devoted to the study of such systems, but no systematic methodology for isolating and quantifying rates \textit{within}
the network has been developed. Similarly, there has been no versatile method proposed to define components of interest within the network.
Studies thus far developed  are
tied to a particular choice of basis and particular network partitioning, a significant limitation that restricts 
their applicability. For example, our earlier NESS rate studies  showed that the rate of producing product 
was determined by the long time scale associated with the  
absorption of the weak incoherent radiation. \cite{hoki2011,Axelrod2018,Axelrod2019} This limitation did correctly identify the rate
determining step, but prevented obtaining rates of processes \textit{within} the network. These issues are resolved in 
this paper, where we introduce a projection operator methodology that provides versatility in defining network ``components", and
hence a methodology for isolating and extracting individual rates within the network.

We address these  issues  by beginning with the quantum Liouville equation
that describes the evolution of the populations and coherences
of  a  generic system. We introduce general projectors in the
Liouville space that define the components of the network
and cast the Liouville dynamics into an equation
of   motion   for  the  component populations by  formally  folding other characteristics such as the
coherences  into  the population equations. This folding and consideration of the time-independent 
{NESS} limit results in a kinetic system of rate equations for arbitrarily defined network
components.

The key focus on 
the NESS due to the significance noted above gives rates that are exact in the
NESS case. The  components comprising  the
resulting  network  are  completely general,  built upon the form  of 
projection  operators defining network partitioning. In particular,
the methodology allows us to define partitions to extract rates \textit{within} the network
and to focus on various different processes and rates within the network.

 This  approach  is  relevant  for  arbitrary
systems,  and  significantly  generalizes  previous  works of
quantum network analysis.~\cite{Cao2009,Wu2012,Roden2016,Liu2019,Yang2020,Jung2020}
 We also provide
insight  into  how  and  when  one  may  measure these rates by
explicitly  perturbing  the  system  out  of its {NESS} and
following the relaxation back to the steady state.



One additional note is in order.
Natural light-induced processes are often experimentally studied via pulsed laser excitation, a light source that is totally alien to natural phenomena.
For example, in the first steps in vision, retinal undergoes {\it cis - trans} isomerization, with the quantum yield of approximately 65\%. 
Pulsed laser experiments imply that this process occurs on a timescale as fast as 60 femtoseconds\cite{Johnson2017}.
However, this rate is of only marginal relevance for isomerization under natural incoherent light.
Rather, what is required is \textit{the rate of the process as it exists within the natural {NESS}}, a focus of this paper.
Indeed, we show below that, as anticipated\cite{Brumer2018} the  {NESS}  rates  are
dramatically  different  from  those  obtained  from a vertical
excitation  processes  typical  in ultrafast pulse excitations.
This  finding  clearly demonstrates the need for theoretical/computational and experimental studies of the
{NESS}  regime  to properly model natural processes.

This  paper  is  organized as follows. Section \ref{thebuild}
demonstrates how
the  Liouville  equation,  in  the  {NESS}  limit,  can  be
rewritten as a kinetic network, using Mori-Zwanzig and Feshbach
projectors						       
\cite{Zwanzig1961,Mori1965,Feshbach1958,Feshbach1962},  allowing
us  to define a partitioning of the network into various constituents  of  interest.
These projectors follow a set of basic rules, admit 
a wide range of possibilities and provide the formal definition
of  the  {NESS} rate matrix. The approach is applied
to  two  models in Sec. \ref{sec:Vsys} based on a minimal model of energy transfer comprised of
three-levels   that   is   analytically   soluble,   and   in Sec. \ref{sec:Brown} to  a
non-equilibrium  spin-boson model. The latter, a minimal model for processes that relate to
isomerization in biophysics, uses a projector defined in
the  nuclear  space,  which is beyond the treatment of previous
methods.

 Section \ref{deviate} then deals with the issue of the validity of the rate
expression away from the NESS, the relation to Markovian vs. non-Markovian dynamics, and
the utili1ty of perturbing the system away from the NESS to obtain the rate by a fitting
procedure. We conclude with final remarks.

\section{Constructing kinetic networks from the Liouville Equation}
\label{thebuild}
\subsection{The construction}

Consider the Liouville equation
\begin{equation}\label{eq:GenQME}
\dot{\hat{\rho}} = \mathcal{L}[\hat{\rho}],
\end{equation}
where the system density matrix is $\hat{\rho}$, the dot denotes the time-derivative, and $\mathcal{L}$ is the Liouvillian.
The Liouvillians are designed to  allow energy but not population flows between the system and the environment.
Furthermore, since our goal is to construct the kinetic system of equations for a steady state,
the Liouvillian is Markovian\cite{Manzano2020-qs} and we assume that it has a single steady state, $\hat\rho_s$,
\begin{equation}\label{eq:NESSQME}
\mathcal{L}[\hat{\rho}_s]=0.
\end{equation}
For such Liouvillians, we propose to construct the general form of a kinetic network~\cite{kreuzer1981nonequilibrium,Fischetti1998}
\begin{equation}\label{eq:GenCME}
\dot{\boldsymbol{p}} = \boldsymbol{k} \boldsymbol{p},
\end{equation}
where $\mathbf{k}$ is a transition rate matrix, and $\mathbf{p}$ is a vector of populations of interest.
The corresponding steady state populations $\boldsymbol{p}_s$ for $N$ network components satisfy the global  balance condition
\begin{equation}\label{eq:NESSCME}
\sum_{n \neq m}^N \left( k_{mn}p_{s,n} - k_{nm}p_{s,m} \right) = 0.
\end{equation}
where $p_{s,n}$ is the $n^{th}$ element of $\bf{p}_s$.

To define rates between arbitrary components, we divide the system into $N$ components and  consider
the population flow between these parts.
Populations of these parts can be defined as $p_n=\tr\{\hat P_n\hat\rho\}$ where $\{\hat P_n\}_{n=1}^N$ form a complete set 
of projectors in the Hilbert space satisfying
\bea
\sum_n\hat P_n &=& 1, \label{eq:HProjP1}\\
\hat P_m\hat P_n &=& \hat P_n\delta_{mn}. \label{eq:HProjP2}
\eea

To partition the Liouville equation, the \emph{Hilbert} space projectors  $\hat{P}_n$ are transformed into \emph{Liouville} space projectors $\Pi_n$ as
\begin{align}\label{eq:LProjP}
\PL_n[\hat{B}] &= \hat\varrho_n\tr\{\hat{P}_n \hat{B}\},
\end{align}
where $\hat{B}$ is any operator, and the operators $\{\hat\varrho_n\}_{n=1}^N$ are partitioned and renormalized components of the steady state density
\bea\label{eq:truncatedSS}
\hat\varrho_n &=& \frac{\hat P_n\hat\rho_s\hat P_n}{\tr\{\hat P_n\hat\rho_s\}}.
\eea
The projectors $\{\PL_n\}_{n=1}^{N}$ do not form a complete set,  but can be completed 
with the addition of the projector $\mathcal{Q}$ on the complementary Liouville space, with 
\begin{align}
\mathcal{Q} &= 1 - \sum^N_{n=1} \PL_n.
\end{align}
Applying the  projectors $\{\PL_n\}$ and $\mathcal{Q}$ to Eq.~(\ref{eq:GenQME})
and using the relation $\tr\{\hat P_m\PL_m[\hat\rho]\}=p_m$, gives the following set of equations involving the populations $p_n$:
\begin{align}
\label{eq:sys_of_eqnsP}
\dot{p}_m &= \sum^N_{n=1}\tr\{\hat{P}_m\mathcal{L}[\hat{\varrho}_n]\} p_n + \tr\{\hat P_m\mathcal{L}\mathcal{Q}[\rho]\}, \\
\label{eq:sys_of_eqnsQ}
\mathcal{Q}[\dot{\hat{\rho}}] &= \sum^N_{n=1}\mathcal{Q}\mathcal{L}[\hat{\varrho}_n] p_n + \mathcal{Q}\mathcal{L}\mathcal{Q}[\hat{\rho}].
\end{align}
Reduced dynamical equations containing only populations are obtained by integrating Eq.~(\ref{eq:sys_of_eqnsQ})
\bea\label{eq:Q}
\mathcal{Q}[\hat{\rho}(t)] &=&
\e{\mathcal{Q}\mathcal{L}\mathcal{Q}t}\mathcal{Q}[\hat{\rho}(0)] \nonumber\\&&
+ \sum^N_{n=1} \int_0^t \mathrm{d}\tau \e{\mathcal{Q}\mathcal{L}\mathcal{Q}(t-\tau)}\mathcal{Q}\mathcal{L}[\hat{\varrho}_n] p_n (\tau).
\eea
and substituting the result into Eq.~(\ref{eq:sys_of_eqnsP}) to give
\bea\notag
\dot{p}_m &=&
  \sum^N_{n=1}\tr\{\hat P_m\mathcal{L}[\hat{\varrho}_n]\} p_n
+ \tr\{\hat P_m\mathcal{L}\e{\mathcal{Q}\mathcal{L}\mathcal{Q}t}\mathcal{Q}[\hat{\rho}(0)]\}\\
&&+ \sum^N_{n=1} \int_0^t \mathrm{d}\tau \tr\{\hat P_m\mathcal{L}\e{\mathcal{Q}\mathcal{L}\mathcal{Q}(t-\tau)}\mathcal{Q}\mathcal{L}[\hat{\varrho}_n]\} p_n (\tau).~\quad\label{eq:red_eqnsp}
\eea
While Eq.~(\ref{eq:red_eqnsp})  is exact, it is not yet in the form of Eq.~(\ref{eq:GenCME}).
To arrive at the kinetic equation form, we focus on the steady state by
substituting $\hat\rho(0)$ by $\hat\rho_s$ and $\boldsymbol p$ by $\boldsymbol p_s$ in Eq.~(\ref{eq:red_eqnsp}).
Then, $\boldsymbol p_s$ becomes time-independent and the time integral can be done analytically.
Furthermore, Eq.~(\ref{eq:sys_of_eqnsQ}) for the steady state becomes $\mathcal{Q}[\hat{\rho}_s]=-\sum_{n=1}^{N}(\mathcal{Q}\mathcal{L}\mathcal{Q})^{-1}\mathcal{Q}\mathcal{L}[\hat\varrho_n]p_{s,n}$, which brings Eq.~(\ref{eq:red_eqnsp}) to the desired form
\begin{align}\label{eq:red_eqnsp_SS}
\dot{\boldsymbol p}_s &=
\boldsymbol k \boldsymbol p_s = \boldsymbol 0,
\end{align}
where $\boldsymbol k$ is a time-independent transition rate matrix whose elements are
\begin{align}\label{eq:ourrate}
k_{mn} &=
\tr\{\hat P_m(1-\mathcal{L} (\mathcal{Q}\mathcal{L}\mathcal{Q})^{-1} \mathcal{Q})\mathcal{L}[\hat{\varrho}_n]\}.
\end{align}
This rate definition satisfies global balance in the steady state (Eq.~(\ref{eq:NESSCME}), i.e., $\boldsymbol{k}\boldsymbol{p}_s=\boldsymbol{0}$).
Also, the construction of the rate matrix does not depend on the particular type of steady state, and can equally be applied to equilibrium or non-equilibrium  circumstances.

The rates defined in Eq. (\ref{eq:ourrate}) contain no approximations and are valid at the {NESS}.
They {\it contain all information relevant to the transfer of population from one part to another, including the effects of the coherences}. 

Note that the choice of the Hilbert space projectors used in Eq.~(\ref{eq:LProjP}) is not limited to any particular
form, as long as they satisfy Eqs.~(\ref{eq:HProjP1}-\ref{eq:HProjP2}).
While we limit the presentation to cases where the projectors are defined either in the nuclear or electronic
subspaces of the system, the projector form is flexible and can be chosen to suit any problem of interest.

Note that there is no need for dynamics simulations to obtain the steady state rates using Eq. (\ref{eq:ourrate}).
Rather, we can solve equation $\mathcal{L}\hat\rho_s = 0$  to obtain $\hat\rho_s$ and then construct the inverse
of the superoperator $\mathcal{Q}\mathcal{L}\mathcal{Q}$.
For small systems this is a simple task, which can be achieved by solving the system of Eqs.~(\ref{eq:sys_of_eqnsP}-\ref{eq:sys_of_eqnsQ}) in the steady state limit, as done in Appendix \ref{app:rates_lineq}.
However, for large systems this is nontrivial, since the size of the Liouvillian scales quadratically with the basis size.
Approaches that avoid explicit construction of the full Liouvillian, such as the stabilized conjugate gradient method\cite{BCG1976},
or the iterative approach introduced in Ref. \citenum{Axelrod2018}, are more efficient for large systems.

Finally, note that the resultant kinetic network Eq. (\ref{eq:red_eqnsp_SS}) applies explicitly at the NESS.
However, deviations from the steady state, and the associated return to the NESS are also of interest, as
discussed in Sec. {\ref{deviate}, where non-Markovian and Markovian contributions to  Eq.~(\ref{eq:red_eqnsp}) from the complimentary space
are analyzed in detail.

To examine rates in various systems of interest, and expose the utility of the choice of projectors, we consider
two sample systems below.

\section{V-system}
\label{sec:Vsys}

Consider first the V-system, a three level
system that has been the subject of great interest in quantum optics \cite{Agarwal2013}, coherence phenomena \cite{Ficek2004},
population trapping \cite{Scully1997} and qubit-qubit interactions and biophysical population dynamics\cite{Tscherbul2018-rs},
and as a minimal energy transfer donor-acceptor model\cite{Jung2020}.
 In this case 
(Fig. \ref{fig:Vsys}) the Hamiltonian is of the form
\begin{equation}
\hat{H}_s = \epsilon_g |g\rangle\langle g| + \sum^2_{k= 1} \epsilon_k |k\rangle\langle k| + J (|1\rangle\langle 2| + |2\rangle\langle 1| )
\label{eq:HS}
\end{equation}
The system contains a ground state $|g\rangle$ and two excited states $|1\rangle$ and $|2\rangle$
coupled with  strength $J$. The system is connected  to hot and cold baths.
The hot bath excites the ground state to  the excited states as
\begin{equation}
\label{eq:LHOT}
\mathcal{L}_{H}[\hat{\rho}] = 2\sum^2_{k=1}\Gamma_{H_k}
 \left(|1\rangle\langle g| \hat{\rho} |g\rangle\langle 1| -\frac{1}{2}[|g\rangle\langle g|, \hat{\rho}]_+\right).
\end{equation}
The cold bath is coupled with the system in two ways:
(i) with a term that is responsible for de-excitation from the excited manifold back to the ground state
\begin{equation}
\label{eq:LCOLD}
\mathcal{L}_{C}[\hat{\rho}] = 2\sum^2_{k=1}\Gamma_{C_k}
 \left(|g\rangle\langle k| \hat{\rho} |k\rangle\langle g| -\frac{1}{2}[|k\rangle\langle k|, \hat{\rho}]_+\right),
\end{equation}
and (ii) with terms responsible for relaxation and dephasing within the excited manifold
\begin{eqnarray}
\label{eq:LD}
\mathcal{L}_{D}[\hat{\rho}] &=&
2\Gamma_{D_f} \left(| 2\rangle\langle 1| \hat{\rho} |1\rangle\langle 2|  -\frac{1}{2}[|2\rangle\langle 2|, \hat{\rho}]_+\right) \nonumber\\ && +  2\Gamma_{D_b} \left(| 1\rangle\langle 2| \hat{\rho} |2\rangle\langle 1|  -\frac{1}{2}[|1\rangle\langle 1|, \hat{\rho}]_+\right).
\end{eqnarray}
Here $[|i\rangle \langle i|,\hat{\rho}]_+ = |i\rangle \langle i|\hat\rho + \hat{\rho}|i\rangle\langle i|$.
The total master equation for the system reduced density matrix $\hat{\rho}$ is
\begin{equation}
\dot{\hat{\rho}} = -i[\hat{H}_S,\hat{\rho}] +\mathcal{L}_{\textrm{H}}[\hat{\rho}] + \mathcal{L}_{\textrm{C}}[\hat{\rho}] + \mathcal{L}_D[\hat{\rho}].
\label{eq:VsysME}
\end{equation}

To illustrate the versatility of our approach, we consider the V-system with two different choices of partitions,
where ground and first excited states are either 
ungrouped  (the standard case in the literature) or grouped together, while the second excited state always remains in its own group (Fig. \ref{fig:Vsys}).
Unless stated otherwise, the numerical values of the parameters are set to the values in Table~\ref{tab:Vsystem_param}.

\begin{table*}
\centering
\caption{Parameters employed in the V-system model in atomic units.}
\newcommand{\size}{0.3cm}
\begin{tabular}{@{\hspace{\size}}c@{\hspace{\size}}c@{\hspace{\size}}c@{\hspace{\size}}c@{\hspace{\size}}c@{\hspace{\size}}c@{\hspace{\size}}c@{\hspace{\size}}c@{\hspace{\size}}c@{\hspace{\size}}c}\hline
$\epsilon_g$ & $\epsilon_1$ & $\epsilon_2$ & $J$ & $\Gamma_{H_1}$ & $\Gamma_{C_1}$ & $\Gamma_{C_2},\Gamma_{D_f}$ & $\Gamma_{H_2},\Gamma_{D_b}$ \\\hline
$0$ & $0.02$ & $0.012$ & $2\cdot10^{-5}$ & $1.5\cdot10^{-6}$ & $4.5\cdot10^{-6}$ &  $10^{-9}$ & $0$ \\\hline
\end{tabular}
\label{tab:Vsystem_param}
\end{table*}

\subsection{Standard V-System}

Using Eq. (\ref{eq:VsysME}) and choosing the Hilbert space projectors $\{\hat{P}_i=|i\rangle\langle i|;i=g,1,2\}$ (in this case $\hat\varrho_i=\hat P_i$) 
we obtain the equations of motion for the populations in the steady state as, where atomic units are used throughout,
\begin{align}
\label{eq:EoMsg}
\dot{\rho}_{gg}
&= -2(\Gamma_{H_1}+\Gamma_{H_2}) \rho_{gg} +2\Gamma_{C_1} \rho_{11} +2\Gamma_{C_2} \rho_{22},
\\
\label{eq:EoMs1}
\dot{\rho}_{11}
&=
 2\Gamma_{H_1} \rho_{gg}
-2\left( \beta +\Gamma_{C_1} +\Gamma_{D_f} \right) \rho_{11}
+2\left( \beta +\Gamma_{D_b} \right) \rho_{22},
\\
\label{eq:EoMs2}
\dot{\rho}_{22}
&=
 2\Gamma_{H_2} \rho_{gg}
+2\left( \beta +\Gamma_{D_f} \right) \rho_{11}
-2\left( \beta +\Gamma_{C_2} +\Gamma_{D_b} \right) \rho_{22},
\end{align}
and where we used the definitions
\begin{align}
\label{definitions}
\beta    &= \frac{J^2\Gamma^*}{(\Gamma^*)^2 +\Delta^2}, \\
\Gamma^* &= \Gamma_{C_1} +\Gamma_{C_2} +\Gamma_{D_f} +\Gamma_{D_b}, \\
\Delta   &= \varepsilon_2-\varepsilon_1.
\end{align}
Here, $\beta$ represents the component of the rate induced by the coherence.
Further details regarding the derivation of these equations are provided in Appendix \ref{app:proof_V-syst_ungrouped}.

\begin{figure}[h!]
\centering
\includegraphics[width=0.5\textwidth]{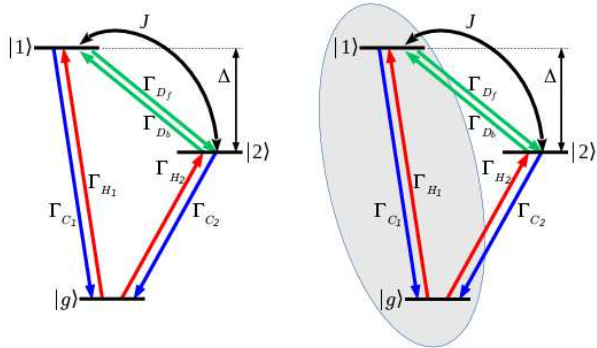} 
\caption{Left: Depiction of the  V-system. The hot bath transition is in red, cold bath transition are in blue, and the dephasing bath is in green. $\Delta$ denotes the excited state splitting. Right: The grouped V-system. The states enclosed in light blue depict the composite state $A$, which is comprised of the ground state and the first excited state.}
\protect\label{fig:Vsys}
\end{figure}

Given  Eqs. (\ref{eq:EoMsg}) - (\ref{eq:EoMs2}) the rates between state 1 and state 2 are seen to be
\begin{align}
k_{21} &= 2\left( \beta +\Gamma_{D_f} \right), \\
k_{12} &= 2\left( \beta +\Gamma_{D_b} \right).
\end{align}
The forward rate $k_{21}$ drives the population from $\rho_{11}$ to $\rho_{22}$, while the backward rate $k_{12}$ transfers the population in the
 reverse direction.
Both rates contain two terms originating from distinct mechanisms:
(1) The term $\beta$ quantifies population transfer through the coherences $\rho_{12}$.
Its presence is a direct result of rewriting the equations of motion of the system entirely in terms of populations, i.e., from making the populations implicitly dependent on the coherences.
(2) The terms $2\Gamma_{D_f}$ or $2\Gamma_{D_b}$ give  population transfer through the phonon bath.
Note that if one regards the coherences
as quantum,  then the first term is a quantum contribution and the second is classical.
Interestingly, both $k_{12}$ and $k_{21}$ show the same dependence on $\beta$; that is, they only differ through the effect of the coupling to the bath.
Hence the coherence term does not favor the forward or backward rate, nor does the sign of the level spacing $\Delta$,
since it enters as $\Delta^2$. Rather, it is the coupling to the bath that determines the directionality of the population flow.

Note that defining the projectors as $\{\hat{P}_i=|i\rangle\langle i|;i=g,1,2\}$ has 
successfully isolated the specific rate between states 1 and 2;
indirect rates of population transfer between  states 1 and 2, such as via $\Gamma_{C1}+ \Gamma_{H2}$, do not participate.

Although the V-system is a simple model, examining the parameter dependence of the system and bath couplings is warranted
to suggest dependences in large natural systems.
When the bath-induced coupling between state
$\ket{1}$ and $\ket{2}$ is zero, i.e., $\Gamma_{D_f}=\Gamma_{D_b}=0$,
the larger the $\Gamma^*$ or $\Delta$, the smaller the $k_{12}=k_{21}$ rate. 
For degenerate states, $\Delta=0$, the transfer rate decreases as $\Gamma^{*}$ increases, due to destructive effects of the bath on the coherences.
When $J=0$  population transfer is through $\mathcal{L}_D$, and
the rate of excited state transfer resembles a classical rate\cite{Thoss2001,Xu2016} in which the transfer is mediated through the bath instead 
of via the coherences  between the excited states.
Limiting cases such as those  described above demonstrate that our rate definition
agrees with previous analysis of quantum networks and the role that coherences play in them.\cite{Liu2019,Cao2009,Wu2012,Engel2007,Manzano2013,Chuang2020,Jung2020}

Additional parameters of interest include those that induce asymmetry\cite{Jung2020}.
As noted above, the nondegenerate case where
$\Delta\neq0$ [from Eqs. (\ref{eq:EoMs1}) and (\ref{eq:EoMs2})], is  detrimental to the rate. Asymmetry can also be introduced by having different parameters for the two excited states.
For example, we can design a ``circular'' flow by making $\Gamma_{D_f}\gg\Gamma_{D_b}$, and $\Gamma_{H_1}\gg\Gamma_{H_2}$ so that
(i) the channel using $\Gamma_{H_1}/\Gamma_{C_1}$ pumps population from ground to first excited state;
(ii) the channel using $\Gamma_{D_f}/\Gamma_{D_b}$ transfers population from the first state to the second excited state; and
(iii) the population is dumped from the second excited state to the ground state using the $\Gamma_{H_2}/\Gamma_{C_2}$ channel.
Hence,  increasing $\Gamma_{D_f}$ and $\Gamma_{C_2}$ accelerates the population decay from state $1$ to the ground state through state $2$.

Further details regarding the interdependence of $\beta$ on $\Delta$ and the bath parameters is provided in
Fig. \ref{fig:bet_gamaC2_gamaDf_Delta}. The value of $\beta$ is seen to behave monotonically at lower $\Gamma=\Gamma_{C_2}=\Gamma_{D_f}$ values, with
larger $\Delta$ leading to smaller $\beta$ and with little dependence on $\Gamma$ for fixed $\Delta$.
The functional form becomes more complex at higher $\Gamma$, displaying
behavior similar to that of environmentally assisted transport\cite{Plenio2008-rw,Rebentrost2009-do,Cao2009}, with a peak in 
$\beta$ with increasing $\Gamma$ at larger $\Delta$.  
For example, for $\Delta = 0.1$ a.u. increasing $\Gamma_{C_2} = \Gamma_{D_f}$ from $10^{-6}$ a.u. to $0.1$ a.u. results in an increase
in $\beta$ by almost two orders of magnitude.

\begin{figure}[h!]
\centering
\includegraphics[width=0.5\textwidth]{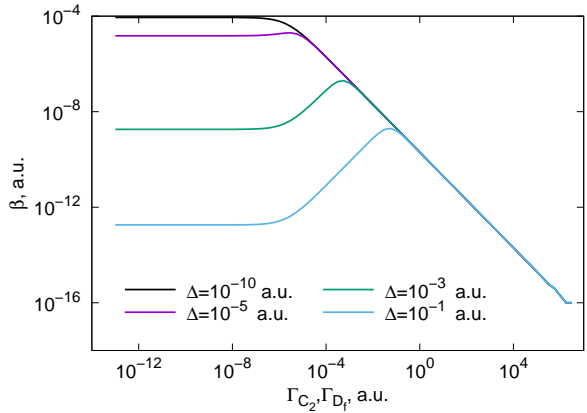} 
\caption{Variation of the coherence component $\beta$ of the rate with respect to the parameters $\Delta$ and $\Gamma_{C_2}=\Gamma_{D_f}$ in the 
V-system.}
\protect\label{fig:bet_gamaC2_gamaDf_Delta}
\end{figure}

\subsection{Grouped V-system model}

As an example of the flexibility of the methodology, consider the case where $|g\rangle$ and $|1\rangle$ are grouped together and where  the population transfer between this group and state $2$
is of interest (Fig. \ref{fig:Vsys}). This constitutes a totally different definition of the components within the network than that considered above.
To do so  we redefine the projectors  to construct a network for two groups rather than three via 
the projectors $\hat P_A=|g\rangle\langle g| + |1\rangle\langle1|$ and $\hat P_2=|2 \rangle\langle 2|$.  
Doing so describes a network   in terms of the populations $p_{A}= \rho_{gg} + \rho_{11}$ and $p_{2}$, with the
{NESS} rates between the two groups in this network given as  (see Appendix~\ref{app:proof_V-syst_grouped} for a derivation)
\begin{align}
\label{eq:k2A}
k_{2A} &=
2 \frac{
 (\Gamma_{H_1}+\Gamma_{H_2}) (\beta+\Gamma_{D_f})
+ \Gamma_{H_2} \Gamma_{C_1}
}{ r(\beta+\Gamma_{D_f}) + \Gamma_{H_1} + (1-r) \Gamma_{H_2} + \Gamma_{C_1} }, \\
\label{eq:kA2}
k_{A2} &=
2 \frac{
  \left\{ \splitfrac{ (\beta+\Gamma_{D_f}+\Gamma_{H_1}+\Gamma_{C_1}) \Gamma_{C_2} }{
              + (\beta+\Gamma_{D_b}) ( \Gamma_{H_1} + \Gamma_{H_2} + \Gamma_{C_1} )} \right\} 
}{ r(\beta+\Gamma_{D_f}) + \Gamma_{H_1} + (1-r) \Gamma_{H_2} + \Gamma_{C_1} },
\end{align}
where  $r$ is a ratio of the ground state population within group $A$ in the non-equilibrium steady state [here denoted by superscript
${s}$], 
\begin{equation}
r=\frac{\rho_{gg}^{(s)}}{\rho_{gg}^{(s)}+\rho_{11}^{(s)}}.
\end{equation}
a quantity that  generally depends on all parameters.
These rates completely account for the coherence $\rho_{12}$ and the internal state of group $A$, described by $(1-r) \rho_{gg} - r\rho_{11}$ 
(see Appendix~\ref{app:proof_V-syst_grouped}) in the steady state.  The parameter dependence of the rates is totally different than the 
three level V-system discussed above, and is too complicated 
to allow us to assess conditions under which $k_{2A} > k_{A2}$, driving population onto state 2, or $k_{2A} < k_{A2}$, the reverse.

The key difference between the rate in the grouped model compared to that in  the standard V-system is the role played by the hot bath.
In the latter case, the {NESS} rates between excited states do not depend on $\mathcal{L}_{H}$, whereas rates in the grouped model
are directly influenced by the hot bath.
For example, the forward rate from group $A$ to state $2$  in the latter case is zero if $\mathcal{L}_{H}=0$.

The significance of the hot bath in the grouped model can be understood by noting that forward population transfer can occur via two mechanisms:
(1) population transfer from state $g$ to state $2$ through the term $\Gamma_{H_2}$, and
(2) population transfer from state $1$ to state $2$ through the term $(\beta+\Gamma_{D_f})$.
Both mechanisms, and hence the rate to state $2$, involve the hot bath, the first explicitly and the second implicitly since it requires 
state 1 to be populated.  
The interplay of $k_{2A}$ and the external parameters $\Gamma_{H_1},\Gamma_{H_2}$ is far from trivial and is shown in Fig. \ref{fig:k2A_gamaH1}a.
However, it is clear that 
for $\Gamma_{H_2}=0$ and $\Gamma_{H_1}<10^{-4}$, the forward rate $k_{2A}$ increases near-linearly as a function of $\Gamma_{H_1}$.

As for the backward rate $k_{A2}$, it also depends on $\mathcal{L}_H$ but this rate is non zero even when $\mathcal{L}_H=0$ as can be seen in
Fig.~\ref{fig:k2A_gamaH1}b and understood from  \eq{eq:kA2} by setting $\Gamma_{H_1}=\Gamma_{H_2}=0$ and using the fact that $r=1$ in this limit:
\begin{align}
\lim_{\Gamma_{H_1},\Gamma_{H_2}\to0} k_{A2} &=
2 \left( \Gamma_{C_2} + \frac{ (\beta+\Gamma_{D_b}) \Gamma_{C_1} }{ \beta + \Gamma_{D_f} + \Gamma_{C_1} } \right).
\end{align}
Two processes are clearly seen in this last equation: direct population transfer from state $2$ to the ground state through $\Gamma_{C_2}$, 
and the two-step population transfer going through state $1$.

\begin{figure}[h!]
\centering
\begin{tabular}{rl}
a)\vspace{-0.5cm}&\\&\hspace{-0.5cm}\includegraphics[width=0.5\textwidth]{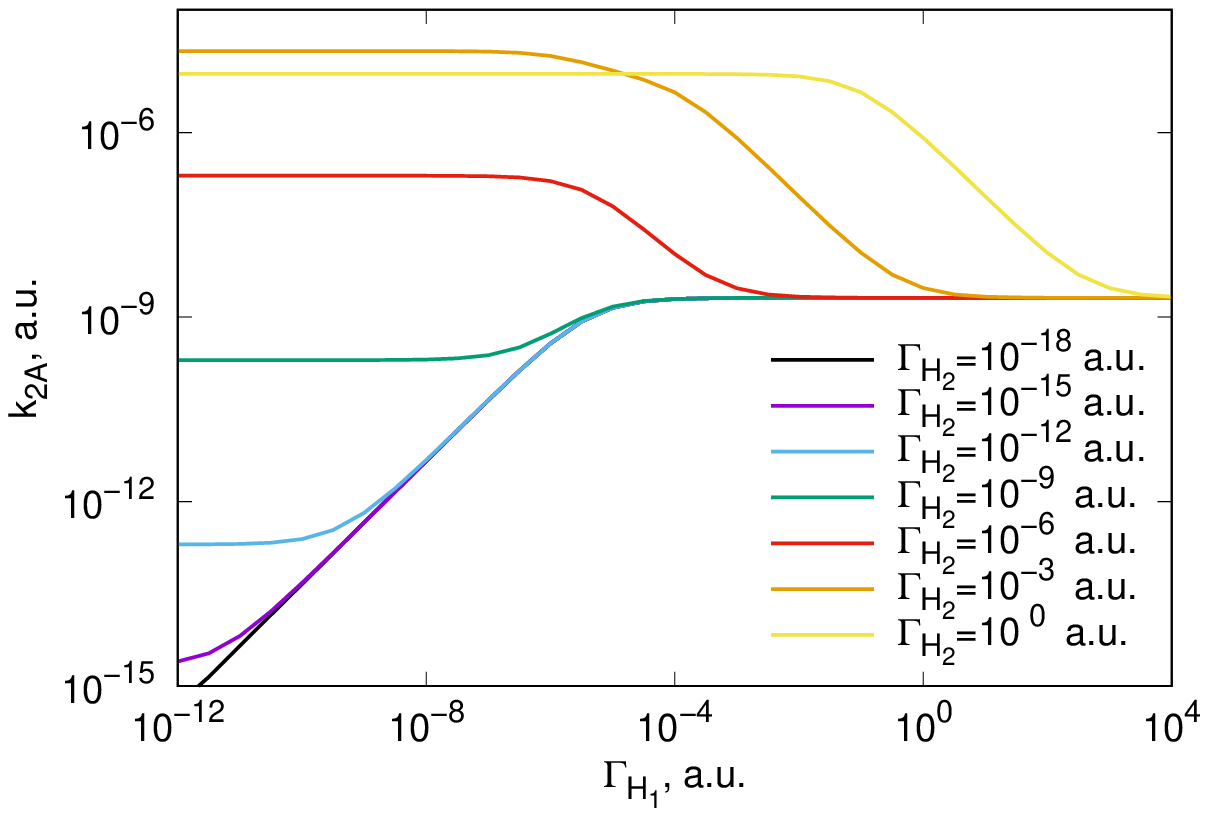} \\
b)\vspace{-0.5cm}&\\&\hspace{-0.5cm}\includegraphics[width=0.5\textwidth]{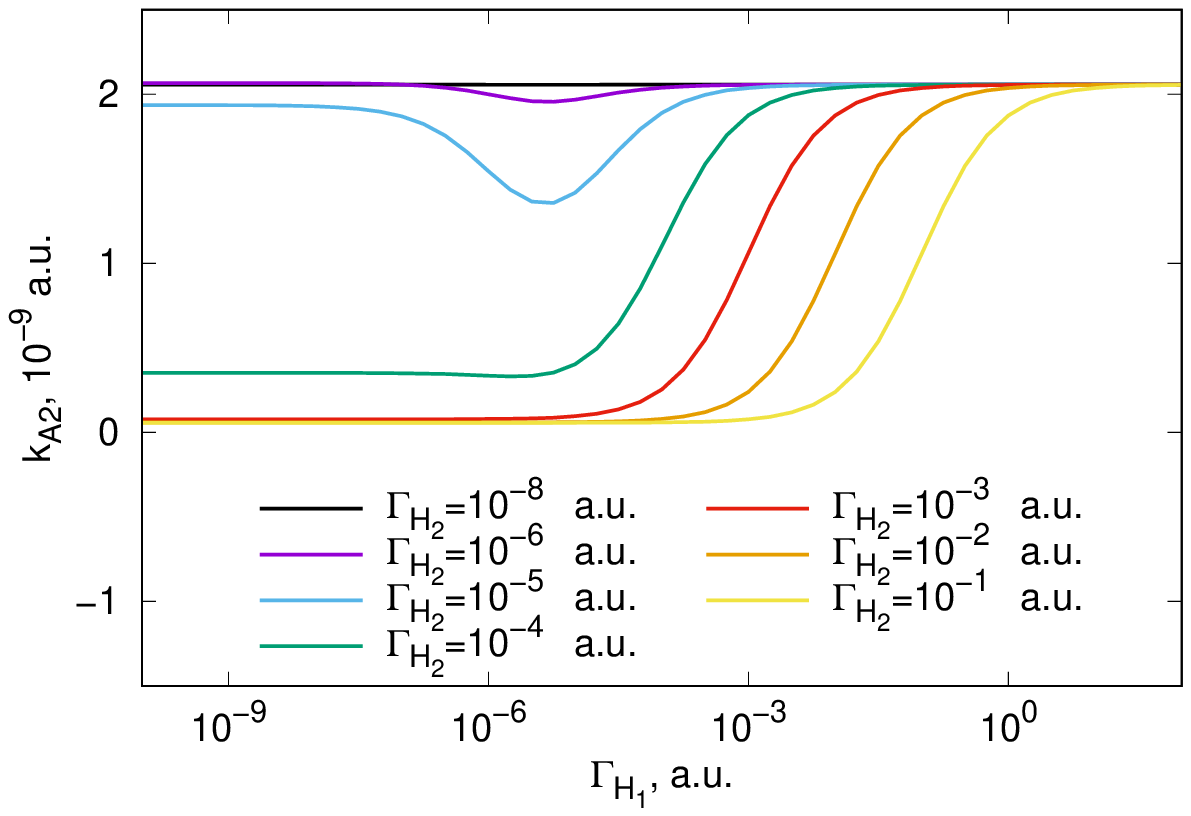}
\end{tabular}
\caption{Variation of (a) the forward rate $k_{2A}$, and (b) the backward rate with respect to the parameters $\Gamma_{H_1}$ and $\Gamma_{H_2}$ in the grouped V-system model.}
\protect\label{fig:k2A_gamaH1}
\end{figure}


An interesting feature of both rates, which can be observed in Figs.~\ref{fig:k2A_gamaH1}a and \ref{fig:k2A_gamaH1}b, is that they
are both bound in the limit of large $\Gamma_{H_1}$ or $\Gamma_{H_2}$.
This effect can be simply exposed by taking the appropriate limits of Eqs.~(\ref{eq:k2A}-\ref{eq:kA2}) and using the fact that
$r=0$ in this limit:
\begin{align}
\lim_{\Gamma_{H_1}\to\infty} k_{2A} &= 2 ( \beta+\Gamma_{D_f} ), \\
\lim_{\Gamma_{H_2}\to\infty} k_{2A} &= 2 ( \beta+\Gamma_{D_f}+\Gamma_{C_1} ), \\
\lim_{\Gamma_{H_1}\to\infty} k_{A2} &= 2 ( \beta+\Gamma_{D_b}+\Gamma_{C_2} ), \\
\lim_{\Gamma_{H_2}\to\infty} k_{A2} &= 2 ( \beta+\Gamma_{D_b} ).
\end{align}

A comparison of these rates with  Fig. \ref{fig:Vsys} shows how taking the limit of specific large radiative pumping $\Gamma_{H1}$ or $\Gamma_{H2}$
highlights specific pathways for population transfer in the grouped V-system. 
However, even for a strong external perturbations the maximal rate is limited by the cold bath and by the system itself.

In typical natural systems the radiative pumping rate is very small, in which case $k_{2A}$ is proportional to $\Gamma_{H_1} + \Gamma_{H_2}$, which is
then the rate determining step.


\section{Nonequilibrium spin-boson model}
\label{sec:Brown}

As a second example consider now a more general case of a composite system that contains both nuclear and electronic degrees of freedom coupled to two baths.
The system chosen is the spin-boson model\cite{Banerjee2018}, which provides input into such important processes as \textit{cis-trans} isomerization in the first steps
in vision\cite{Hahn2000-eo} and proposed photoswitches\cite{Gonzalez2020}.

The system Hamiltonian in the diabatic electronic basis is given by
\begin{eqnarray}
\hat{H}_S &=&  \sum^2_{k=1} \left[\left(-\frac{\Omega_k}{2}\frac{\partial^2}{\partial \hat{q}^2} + \frac{\Omega_k}{2}(\hat{q} -q_k)^2 + \epsilon_k \right) |k\rangle \langle k| \right] \nonumber \\
&& +  \lambda  \left( |1\rangle \langle 2| + |2\rangle \langle 1| \right),
\label{eq:Brown_Ham}
\end{eqnarray}
where $k$ indexes the electronic state (either $1$ or $2$), $\hat{q}$ is the vibrational coordinate, $\Omega_k$ is the frequency of the oscillator in the $k^{th}$ electronic state, and $q_k$ and $\epsilon_k$ are the horizontal and vertical displacements respectively, and the system is subjected to incident incoherent light.  
Figure \ref{fig:Brown} provides a  representative example of the system considered.

The diabats are coupled to a common cold phonon bath (ph) and to a hot thermal bath (rad) used to mimic incoherent radiation-induced transitions 
between the two diabats.
Both baths are given in the Lindblad form. It assumes that the behavior is  Markovian, which is exact in 
the case of  the NESS.
The master equation for the system excited by incoherent radiation is then given as
\begin{equation}
\dot{\hat{\rho}} = -i[\hat{H}_S,\hat{\rho}] + \sum_{\nu=\{\textrm{rad,ph1,ph2}\}} \mathcal{L}_{\nu}[\hat{\rho}] ,
\end{equation}
with
\begin{eqnarray}
\mathcal{L}_{\nu}[\hat{\rho}] &=& \Gamma_{\nu}(n_{\nu}+1) \left ( \hat{S}_{\nu} \hat{\rho} \hat{S}^{\dagger}_{\nu} -\frac{1}{2}[\hat{S}^{\dagger}_{\nu}\hat{S}_{\nu}, \hat{\rho}]_+ \right) \nonumber \\ &&
+ \Gamma_{\nu}n_{\nu} \left ( \hat{S}^{\dagger}_{\nu} \hat{\rho} \hat{S}_{\nu} -\frac{1}{2}[\hat{S}_{\nu}\hat{S}^{\dagger}_{\nu}, \hat{\rho}]_+ \right).
\end{eqnarray}
Here $\Gamma_{\nu}$ scales the interaction of the bath, $n_{\nu}$ gives the mean number of excitations, and $\hat{S}_{\nu}$ are system operators in the combined nuclear-electronic subspace responsible for the coupling to the baths, defined below.
The mean number of excitation is defined as $n_{\nu}=(\exp(\frac{E_\nu}{k_B T_\nu})-1)^{-1}$ where $k_B$ is the Boltzmann constant, $T_\nu$ is the temperature of the corresponding bath, and $E_\nu$ is the level spacing of the corresponding bath i.e. $E_{ph,k}=\Omega_k$ and $E_{rad}$ is the energy gap between the two diabats at $q=-3$.
The radiation bath directly couples electronic states via dipole coupling, i.e.,
\begin{equation}
\hat{S}_{\textrm{rad}} = |1 \rangle\langle 2|,
\end{equation}
while the phonon bath is coupled to each diabatic state $k$ as
\begin{equation}
\hat{S}_{\textrm{ph},k} = \left( \hat{a}  - \frac{q_k}{\sqrt{2}}  \right)|k \rangle\langle k| .
\end{equation}
Here $\hat{a} = \frac{1}{\sqrt{2}}(\hat{q} + \frac{\partial}{\partial\hat{q}})$ and $\hat{S}_{\textrm{ph},k}$  is the annihilation operator of the oscillator defined in state $k$.
The specific form of dissipators is that of a local quantum master equation~\cite{Takagahara:1978/jpsj/728,Joubert:2015/jcp/134107,Hofer:2017/njp/123037}  such that the chosen model is valid in the weak coupling limit.
\begin{figure}[h!]
\centering
\includegraphics[width=0.5\textwidth]{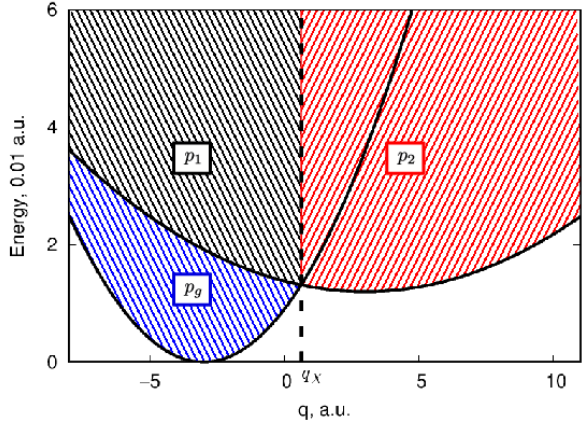} 
\caption{Depiction of the nonequilibrium spin boson model. The dashed line indicates the dividing surface used to partition the nuclear coordinate $q$, defining the position $q_X$.
The hatch patterns depict the partition defined in Eqs.~(\ref{eq:Pg}-\ref{eq:P2}).}
\protect\label{fig:Brown}
\end{figure}

As one partitioning example we identify the rate of interest as that at which population is transferred from the left to the right of a 
dividing surface  (see Fig. \ref{fig:Brown}), chosen to lie at the intersection created between the two diabats  denoted $q_X$ .
Note that this  rate is analogous to that commonly used to study molecular reactions within the reactive flux formalism.
This situation is also reminiscent of the study of \textit{cis-trans} isomerization in retinal \cite{Hahn2000-eo}   in which the crossing of the dividing surface corresponds to the system undergoing isomerization from the \textit{cis}  to the \textit{trans} configuration.
These rates are defined via the Hilbert space projectors that satisfy Eqs.~(\ref{eq:HProjP1}) and (\ref{eq:HProjP2}) as
\begin{eqnarray}
{\hat P}_{L} & = & \int_{-\infty}^{q_X} dq |q\rangle \langle q|, \\
{\hat P}_{R} & = & \int_{q_X}^{\infty} dq |q\rangle \langle q|.
\end{eqnarray}
For a numerical example, parameters (in atomic units)  are taken as follows unless otherwise stated: $\Omega_1 = 2\times10^{-3}$, $\Omega_2 = 4\times10^{-4}$, $\epsilon_2-\epsilon_1 = 0.012$, $q_1=-3.0$, $q_2=3.0$, $\lambda = 2\times10^{-5}$, $\Gamma_{rad}=10^{-6}$, 
$\Gamma_{ph,1}= \Gamma_{ph,2}=10^{-9}$, radiation bath temperature $ T_{rad}=5800K$ and phonon bath temperature $T_{ph}=300K$.

Computations are done by projecting all the operators onto a basis comprised of a direct product of the electronic and nuclear bases,
where the nuclear coordinate basis is chosen as harmonic oscillators centered at $q=0$ a.u. 
(i.e. eigenstates of $\hat{q}^2-{\partial^2}/{\partial \hat{q}^2}$).
We use a large nuclear basis of 400 basis functions to properly represent the projector operators $\hat{P}_R$ and $\hat{P}_L$.
In order to reduce the computational cost, the large basis must be truncated,
while preserving the partition given by the projectors.
To achieve this goal we diagonalize the localized Hamiltonians $\hat{P}_R\hat{H}_s\hat{P}_R$ and $\hat{P}_L\hat{H}_s\hat{P}_L$ and retain only
eigenstates that are lower in energy than $50\Omega_2+\epsilon_2-\epsilon_1$.
We then vectorize the steady state Liouville equation and solve the linear problem of Appendix~\ref{app:rates_lineq} using standard linear
algebra routines.

Processes like this, e.g. molecular isomerization, are often studied with ultrafast laser pulses\cite{Johnson2017}.  For this reason, the rate following a vertical excitation is also of interest.
This is computed by first turning off the photon bath (setting $\Gamma_{\textrm{rad}}$=0) and obtaining the stationary density $\rho_s$.
This density is then excited using a high order perturbation expansion  \cite{Mukamel_Book} to obtain the vertically excited state:

\begin{equation}
    \rho_{\textrm{VE}} = \rho_s + \lim_{n\to \infty} \sum_{k=1}^{\infty} \rho_k
\end{equation}
  Here $\rho_k$ is defined via $k$ nested commutators:
 \begin{eqnarray}
 \rho_k =  \left ( - \frac {i}{h} \right)^k [\mu,[\mu, \cdots,[\mu,\rho_s]\cdots ]]_k ,
 \end{eqnarray}
$\mu = \alpha|2\rangle\langle1|$ + h.c. with  $\alpha = 0.45$  modeling the strength of the dipole in the perturbation expansion.   The resultant excitation is $\sim$ 0.5 eV.  The density is then evolved in time to calculate the transfer rate from one side of the dividing surface to the other.

\begin{figure}[h!]
\centering
\includegraphics[width=0.5\textwidth]{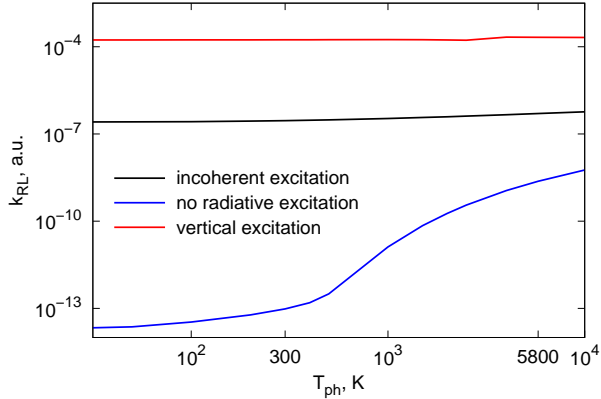}
\caption{Dependence of the forward rate with incoherent excitation (black), vertical excitation (red), and no radiative transition (blue)
 when varying $T_{ph}$.}
\label{fig:Tp_Rates}
\end{figure}

It is well established (e.g., see Refs. \citenum{Shapiro2011} and \citenum{Brumer2018}), but often ignored, that rates of radiatively excited processes depend intimately on the nature of the incident light.  
Figure \ref{fig:Tp_Rates} displays the behavior of (a) the forward {NESS} rate, (b) the rate following the dynamics after a vertical excitation and (c) the rate with no excitation radiative (only phonon bath effects), all with respect to the phonon bath temperature $T_{ph}$.
The three rates are seen to occur with vastly different magnitudes.
The fastest rate, $10^{-4}$ a.u.  $\sim 10^{-2}$ fs$^{-1}$, is that obtained via vertical excitation.  This rate does not depend on the phonon bath temperature since the 
vertical excitation energy of $\sim 0.5$ eV is much larger than phonon bath effects, which only transfer small amounts of population.
Even at the highest temperature shown in Fig.~\ref{fig:Tp_Rates}, excitation due to thermal fluctuations are minuscule compared to that of the vertical excitation.

Figure~\ref{fig:Tp_Rates} shows that the {NESS} rate with incoherent excitation is $\sim3$ orders of magnitude  smaller than that of vertical excitation 
for nearly all temperatures.
The magnitude of this rate is seen to be  nearly independent of the phonon bath for the parameter set of the incoherent light sources  considered here due to the difference in  energy scales.
Significantly, the vertical excitation rate, similar to that achieved in pulsed laser experiments, is orders of magnitude faster than the incoherent
 excitation {NESS} rate, reinforcing the view \cite{Jiang1991,Brumer2018} that rates from pumped laser experiments do not reflect time scales under normal incoherent light.

 Finally, the rate with no radiative excitation is  relevant to, e.g., thermally induced \textit{cis-trans} isomerization of rhodopsin\cite{yanagawa2015}.
 It is, as expected, the slowest.
Here, with system in the dark, the rate of population transfer is independent of the phonon temperature and is very small until $\sim 1000 K$.
At this point (attenuation factor of 0.47) there  is enough thermal energy to overcome the barrier that separates the two minima.
Such an effect is only evident in the absence of radiative contributions.
Even when the temperature of the phonon bath approaches that of the photon bath at $5800K$ there is a large difference between the thermal rate and the radiative rates.
This is mainly due to the fact that, unlike the phonon bath, radiation couples electronic states directly through dipolar coupling, which allows them to transfer population without having to cross a barrier.

\begin{figure}[h!]
\centering
\includegraphics[width=0.5\textwidth]{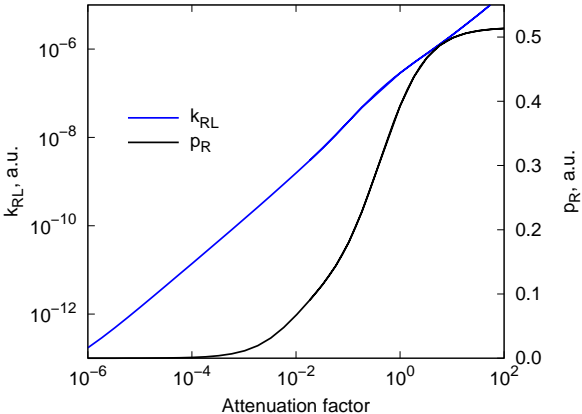} 
\caption{Dependence of the forward rate and the transferred population on the photon attenuation factor $\alpha$.}
\protect\label{fig:Rad_Rate}
\end{figure}

Molecules on earth do not experience the full strength of the solar spectrum due to absorption by the atmosphere. 
Additional attenuation occurs due to artificial or cloud cover, or light absorption by water for undersea plants.
To simulate the  attenuation of the incident incoherent light of the radiative bath, we introduce an attenuation factor $\alpha$ and replace  $n_{rad}$ by the effective quantity $\alpha n_{rad}$.
As seen in Fig.~\ref{fig:Rad_Rate} the dependence of the forward rate on the attenuation factor is linear  on a log-log plot.
This is indicative of power law dependence, and a least square analysis shows  that the rate depends approximately linearly on $n_{rad}$.
Hence, when light is  attenuated by  $\alpha=10^{-2}$, the rate is also reduced  by a factor of $10^{-2}$.  (Realistic attenuation factors in
some photosynthetic systems can be smaller\cite{Chuang2020} than $10^{-7}$.)
This is expected since, under strong attenuation only few photons excite the system to allow for the subsequent population transfer.
Indeed, for a low attenuation factor most of the system is in the first electronic state and populates low energy states localized on the left of the dividing surface (see the transferred population in Fig~\ref{fig:Rad_Rate}).
By unphysically increasing $\alpha$, we can reach a {NESS} rate that is as large as the rate obtained with a vertical excitation.
However, such a large value $\alpha \sim 10^2$  implies a huge unphysical radiation temperature $T\sim3\times10^{5}$ K.

One further note about this scenario is in order.
The comparative behavior of the forward rate and the transferred population, seen in  Fig.~\ref{fig:Rad_Rate}, is also enlightening. In particular,
the transferred population follows the attenuation factor at small $\alpha$. However, the transferred 
population increases dramatically after 
$\alpha \approx 10^{-2}$, reaching a maximum of $\approx 0.5$ after $\alpha \approx 10$. 
This behavior may well be of interest to experiments designing materials operating in, e.g.,  
a solar furnace where temperatures can reach 3000 degrees K. In our case the behavior arises as
follows: When the attenuation factor is small, 
most population is in the lowest energy states, which are located on the left side. As the incoming
energy reaches the energy of the lowest state located on the right side 
(~0.012 a.u., equivalent to a temperature of 3800K), these states become significantly populated. At this point the population 
starts to grow dramatically. 
When the incoming energy is much larger than 0.012 a.u., both left and right states become equally populated,
with the transferred population tending to a limit of 0.5. This type of behavior is clearly system dependent,
and can certainly occur at lower temperatures, depending on system eigenstates.

The general behavior of this system resembles
that of the grouped V-system in displaying a dependence on the external driving field. Such rates, where a weak excitation step is included in the
overall system
definition, correctly identifies  the excitation as the rate determining step in the production of the final state. 
They have been studied in  detail for processes like
energy transfer in LH1\cite{Chuang2020} and the initial steps in vision \cite{hoki2011,Axelrod2018,Axelrod2019}. However, the approach introduced here
allows different choices of projection operators, and hence different partitioning of the network.
This allows us \textit{a focus on the dynamics of the
process within the excited state, a quantity giving insight into the rate of population transfer post excitation that is independent of the 
excitation.} This approach is reminiscent of the (ungrouped) V-system above.

To extract the rate independent of the excitation step we partition the system into three components through the following projection operators
(see Fig. \ref{fig:Brown}):
\begin{eqnarray}
{\hat P}_g & = & {\hat P}_{L} |1\rangle \langle 1|, \label{eq:Pg}\\
{\hat P}_1 & = & {\hat P}_{L} |2\rangle \langle 2|, \label{eq:P1}\\
{\hat P}_2 & = & {\hat P}_{R}. \label{eq:P2}
\end{eqnarray}
Here $\hat{P}_g$ projects onto the lower electronic state on the left side and $\hat{P}_1$ onto the upper electronic state on the left side.
These choices subdivide the system between parts that exchange population through the radiative bath if we assume that most
of the population is located on the left side.
The projector $\hat{P}_2$ is chosen as $\hat{P}_R$, i.e., we do not split the electronic states on the right side because we assume, for the
chosen parameters, that electronic state 1 will be only weakly populated in this region. 
Numerical results are obtained using the same methodology as described above,
and resultant rates are shown in Fig.~\ref{fig:Rad_Rate_3p}.
They clearly display a forward rate, $k_{21}$ which, for $\alpha<10^{-2}$, does not depend on the radiative bath.


This lack of dependence of $k_{21}$ on the attenuation is similar to that observed in the ungrouped V-system case in Sec.~\ref{sec:Vsys},
indicative of the fact that the excitation step has been properly separated from the excited state dynamics within the network.	
This is a significant result, allowing deep insight into the flow of population internal to the network.

\begin{figure}[h!]
\centering
\includegraphics[width=0.5\textwidth]{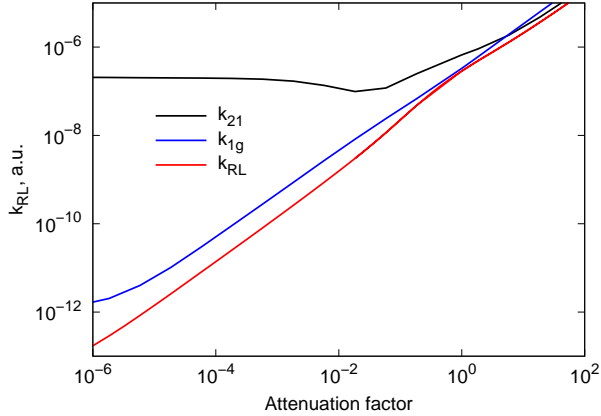} 
\caption{Dependence of the rates $k_{1g}$ and $k_{21}$ on the attenuation factor $\alpha$ in the case of the partition into three 
components. The rate $k_{RL}$ is given for the purpose of comparison.}
\protect\label{fig:Rad_Rate_3p}
\end{figure}

\section{Rates and The Deviation from the NESS}
\label{deviate}

The above results pertain to the all-important NESS region.  When the system is perturbed out of the NESS, the population dynamics
are not expected to be  described  by the rate equation.
However, there are situations where the range of validity of the NESS kinetic equations extends outside the NESS regime.
This is of particular relevance in determining when one can extend
to the NESS, a practice common for equilibrium cases, where equilibrium rates are obtained from the rate of return of a
perturbation back to equilibrium\cite{Yamamoto1960-qm,Miller1983-id}. Interestingly,  as shown below, conditions where this is the case
are intimately related to the role of Markovian vs. non-Markovian contributions to Eq.~(\ref{eq:red_eqnsp}).
Note that these contributions arise from separating the population dynamics from the complementary Liouville space, and do not refer to the
Markovianity or non-Markovianity of the bath.
Here we introduce these conditions and provide an application to the V-system.

Consider then Eq.~(\ref{eq:red_eqnsp}), which can be partitioned into two terms
\begin{align}\label{eq:terms}
\dot{\boldsymbol p} &= \boldsymbol M^{(1)} + \boldsymbol M^{(2)}
\end{align}
where the first term, $\boldsymbol M^{(1)}$ is the contribution from the {NESS} rates defined in Eq.~(\ref{eq:ourrate}),
\begin{align}\label{eq:term1}
M^{(1)}_m &= \sum^N_{n=1} k_{mn} p_n.
\end{align}
This $M^{(1)}$ term generates  Markovian evolution since the generator $k_{mn}$ does not contain memory terms.
The $\boldsymbol M^{(2)}$ term is then the difference $\dot{\boldsymbol p}-\boldsymbol M^{(1)}$:
\begin{widetext}
\begin{align}\label{eq:term2}
M^{(2)}_m &=
\tr\{\hat P_m\mathcal{L}\e{\mathcal{Q}\mathcal{L}\mathcal{Q}t}\mathcal{Q}[\hat{\rho}(0)]\}
+ \sum^N_{n=1}
\int_0^t \mathrm{d}\tau \tr\{\hat P_m\mathcal{L} \e{\mathcal{Q}\mathcal{L}\mathcal{Q}(t-\tau)} \mathcal{Q}\mathcal{L}[\hat{\varrho}_n]\} p_n (\tau)
\nonumber\\&
+ \sum^N_{n=1}
\tr\{\hat P_m\mathcal{L} (\mathcal{Q}\mathcal{L}\mathcal{Q})^{-1} \mathcal{Q}\mathcal{L}[\hat{\varrho}_n]\} p_n.
\end{align}
\end{widetext}

This term contains the time integral memory component
and depends on the initial condition $\mathcal{Q}[\hat{\rho}(0)]$, and hence generates non-Markovian dynamics.
For the derived kinetic equations to be accurate away from NESS, the non-Markovian terms need to be negligible.
The Markovian approximation, which characterizes the NESS, applies if after some transient dynamics of duration $t^{(2)}$ the
terms of $\boldsymbol M^{(2)}$ vanish.
Similarly, we can define the time $t_s$ which is required for the system to reach the steady state.
If $t_s\gg t^{(2)}$, then the populations follow the kinetic equations (or Markovian dynamics) for the time range $t\in[t^{(2)},t_s]$.
The population dynamics during this time range can be used, if desired, to extract the rates by employing a fitting procedure 
without calculating them using Eq.~(\ref{eq:ourrate}).
However, if $t^{(2)}> t_s$, this rate extraction is not possible since the Markovian approximation is not valid for
any time  $t<t_s$. (In either case, however, the NESS rates can, of course, be obtained via Eq. (\ref{eq:ourrate}).)

Consider then the timescales $t^{(1)}$ and $t^{(2)}$, which can be estimated as follows: $t^{(2)}$, is dictated by the decay of $\e{\mathcal{Q}\mathcal{L}\mathcal{Q}t}$.
Hence we define the eigenvalues  of $\mathcal{Q}\mathcal{L}\mathcal{Q}$ and order them as $0>\Re[\kappa_1^{(2)}]>\Re[\kappa_2^{(2)}]>\dots>\Re[\kappa_n^{(2)}]...$.
The slowest decay process $\e{\kappa_1^{(2)}t}$ then determines  $t^{(2)}=1/\Re[\kappa_1^{(2)}]$ a.u.
Once $\boldsymbol M^{(2)}$ becomes negligible, the populations exponentially decay as $\boldsymbol p(t)=\e{\boldsymbol k (t-t^{(2)})}\boldsymbol p(t^{(2)})$.
Hence, this timescale is also governed by an exponential decay  and using  the eigenvalues of the matrix $\boldsymbol{k}$, we order them as $0>\kappa_1^{(1)}>\kappa_2^{(1)}>\kappa_3^{(1)}...$.
The slowest decay process $\e{\kappa_1^{(1)}t}$ determines $t^{(1)}$, which we define as $t^{(1)}=1/\Re[\kappa_1^{(1)}]$ a.u.
With these definitions, the dynamics is Markovian if $t^{(2)}<<t^{(1)}$ and the kinetic rate equations are valid in this time domain.

To gain insight into this analysis, and obtain these time scales for a given system, we could calculate the exact dynamics
and extract the time evolution of both terms.
The exact dynamics would be obtained by exponentiation of the Liouvillian from a perturbed initial state.
The resulting density matrix and its time-derivative would then used to extract ${\boldsymbol p}(t)$ and $\dot{\boldsymbol p}(t)$ at various times.
A detailed study of $t^{(2)},t^{(1)}$ and $t_s$ is the subject of future work, Here we provide one example based on the models introduced above.

\label{Markov}

Consider the case of the V-system in Sec. \ref{sec:Vsys}.  
The parameters given in Table ~\ref{tab:Vsystem_param} give the time scale estimates  $t^{(1)}=4.0 \times 10^8$ and $t^{(2)}=6.6 \times 10^5$ a.u.

Since $t^{(2)}<<t^{(1)}$, we  anticipate that the dynamics after time $t^{(2)}$ will be Markovian and essentially driven by the first term $\boldsymbol M^{(1)}$.
Indeed, this is what is observed in Fig.~\ref{fig:Dyn_Markov}a where $\boldsymbol M^{(2)}<<\boldsymbol M^{(1)}$ after the time $t^{(2)}$.
Since $t^{(2)}$ is an order of magnitude smaller than $t^{(1)}$, the dynamics is expected to follow an exponential decay on the global timescale, whose generator is $\boldsymbol k$.
This is indeed what is observed in Fig.~\ref{fig:Dyn_Markov}b
where the population dynamics obtained from the model given by the steady state rates is compared to the exact dynamics.
This  difference can be quantified by calculating the relative error
\bea\label{eq:error}
\frac{2\int_{t^{(2)}}^{t_f}{\norm{\dot{\boldsymbol p}(t)-\boldsymbol k\boldsymbol p(t)}} \mathrm{d}t}{\int_{t^{(2)}}^{t_f}{\norm{\dot{\boldsymbol p}(t)+\boldsymbol k\boldsymbol p(t)}} \mathrm{d}t},
\eea
where $t_f$ is the final propagation time.
Using a time step $\delta t\approx 7\cdot 10^{4}$ a.u., the obtained relative error is $7.2 \times 10^{-2}$.
We can also extract the rates that would fit best the population curves by minimizing $\int_{t^{(2)}}^{t_f}{\norm{\dot{\boldsymbol p}(t)-\boldsymbol k\boldsymbol p(t)}}^2 \mathrm{d}t$ on the time grid.
The resulting linear equation to solve in order to obtain the fit transition rate matrix is
\bea
\boldsymbol k_{fit} &=& \int_{t^{(2)}}^{t_f} \dot{\boldsymbol p}(t) {\boldsymbol p}^T(t) \mathrm{d}t
\left[ \int_{t^{(2)}}^{t_f} {\boldsymbol p}(t) {\boldsymbol p}^T(t) \mathrm{d}t \right]^{-1},
\eea
where the superscript $T$ denotes the transpose.
Eigenvalues of $\boldsymbol k_{fit}$, $\{0,0,-2.58\cdot 10^{-9}\}$, can be compared to the 
eigenvalues of $\boldsymbol k$, $\{0,-2.58\cdot10^{-9},-1.21\cdot10^{-5}\}$.
Hence, the long timescale is quantitatively recovered from the fitting procedure with an error of less than $10\%$.

\begin{figure}[h!]
\centering
\begin{tabular}{rl}
a)\vspace{-0.5cm}&\\&\hspace{-0.5cm}\includegraphics[width=0.5\textwidth]{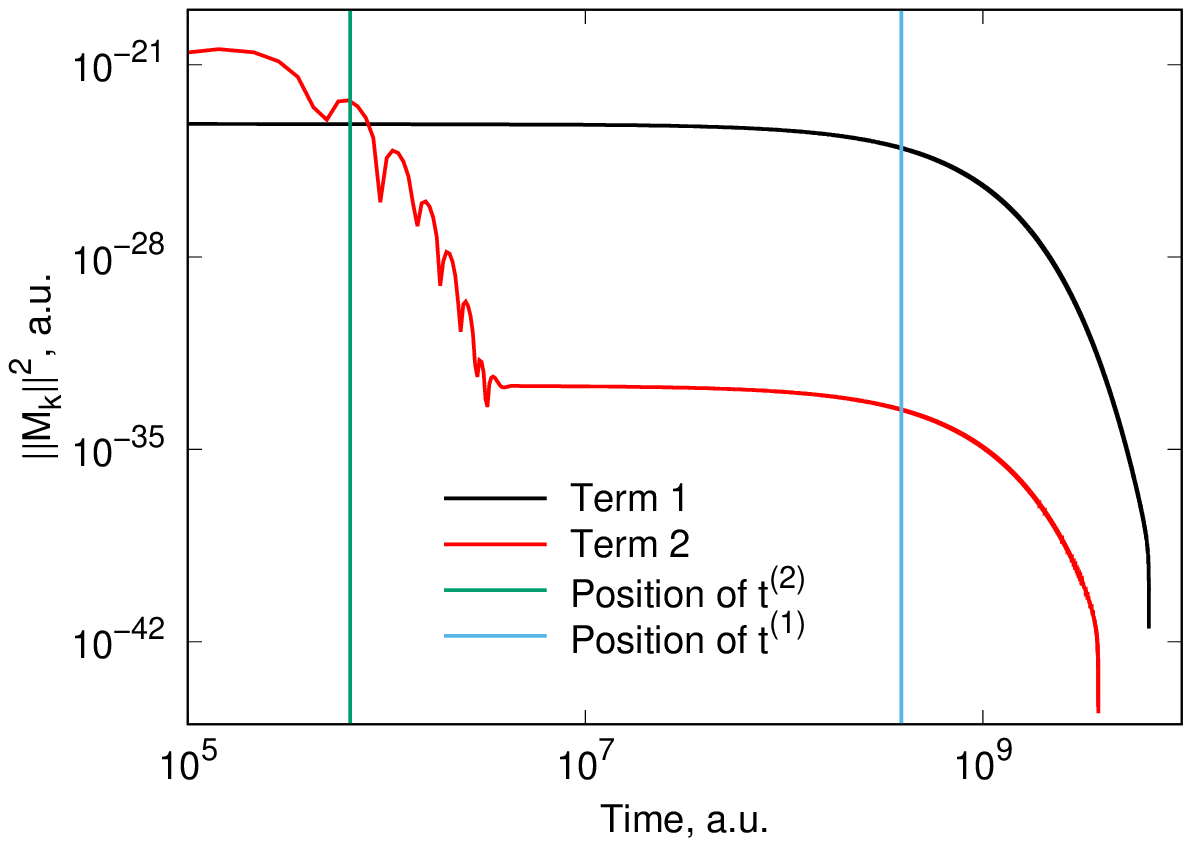} \\
b)\vspace{-0.5cm}&\\&\hspace{-0.5cm}\includegraphics[width=0.5\textwidth]{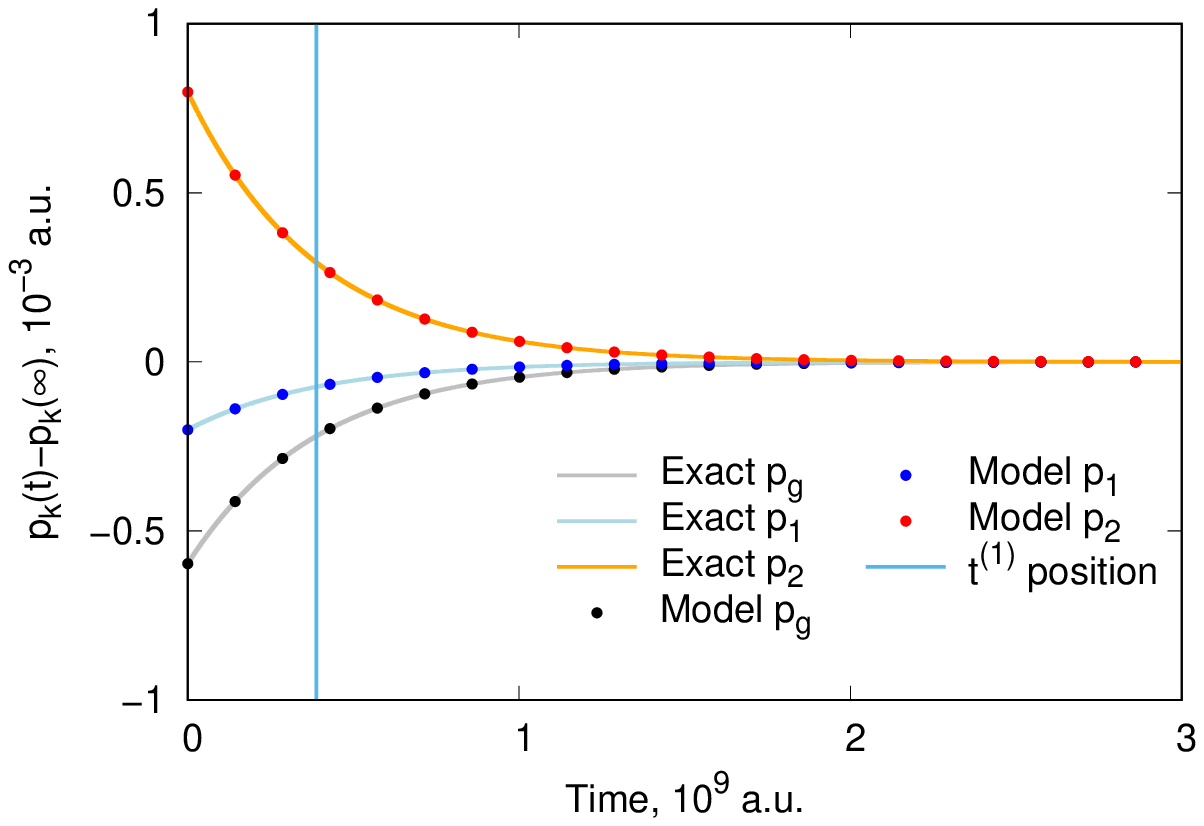}
\end{tabular}
\caption{Time evolution  for the V-system in the Markovian case. (a) $\mathbf{M}^{(1)}$ and $\mathbf{M}^{(2)}$ and
(b) the populations comparing exact propagation to steady state rates .}
\protect\label{fig:Dyn_Markov}
\end{figure}


The alternative situation arises when $J$ [and hence the coherence mediated rate $\beta$ in Eq. (\ref{definitions})] is increased.
This is a signature of the importance of the complementary space, which can be 
quantified by the spectral norm $\norm{[\mathcal{Q}\mathcal{L}\mathcal{Q}]^{-1}}$ in \eq{eq:ourrate}.
This is indeed what we observe when we set $J=0.02$ a.u. and obtain $\Re[\kappa_1^{(2)}]=-3.3 \times 10^{-6}$ a.u.
In this case, the timescale for $\boldsymbol M^{(2)}$ is given by $t^{(2)}=3.0 \times 10^5$ a.u. and the $\boldsymbol M^{(1)}$ timescale
gives $t^{(1)} = 1.4 \times 10^5$.
As a result, $t^{(1)}<t^{(2)}$ and $\boldsymbol M^{(1)}<\boldsymbol M^{(2)}$ for the entire dynamics.
Thus, $\boldsymbol M^{(2)}$ is never negligible and non-Markovianity dominates for all times, as seen in
 Fig.~\ref{fig:Dyn_nonMarkov}a.
The same conclusion is reached by observing the population dynamics in Fig.~\ref{fig:Dyn_nonMarkov}b.
In this case, the relative error defined in \eq{eq:error} is $2$, so that it is impossible to define a time-range
over which rates can be extracted to fit the dynamics.  That is, the NESS rates have to be properly determined from Eq. ({\ref{eq:ourrate}).
This can already be understood by visual inspection of Fig.~\ref{fig:Dyn_nonMarkov}b, where the populations $p_1$ and $p_2$ are oscillating
during the entire time evolution.

\begin{figure}[h!]
\centering
\begin{tabular}{rl}
a)\vspace{-0.5cm}&\\&\hspace{-0.5cm}\includegraphics[width=0.5\textwidth]{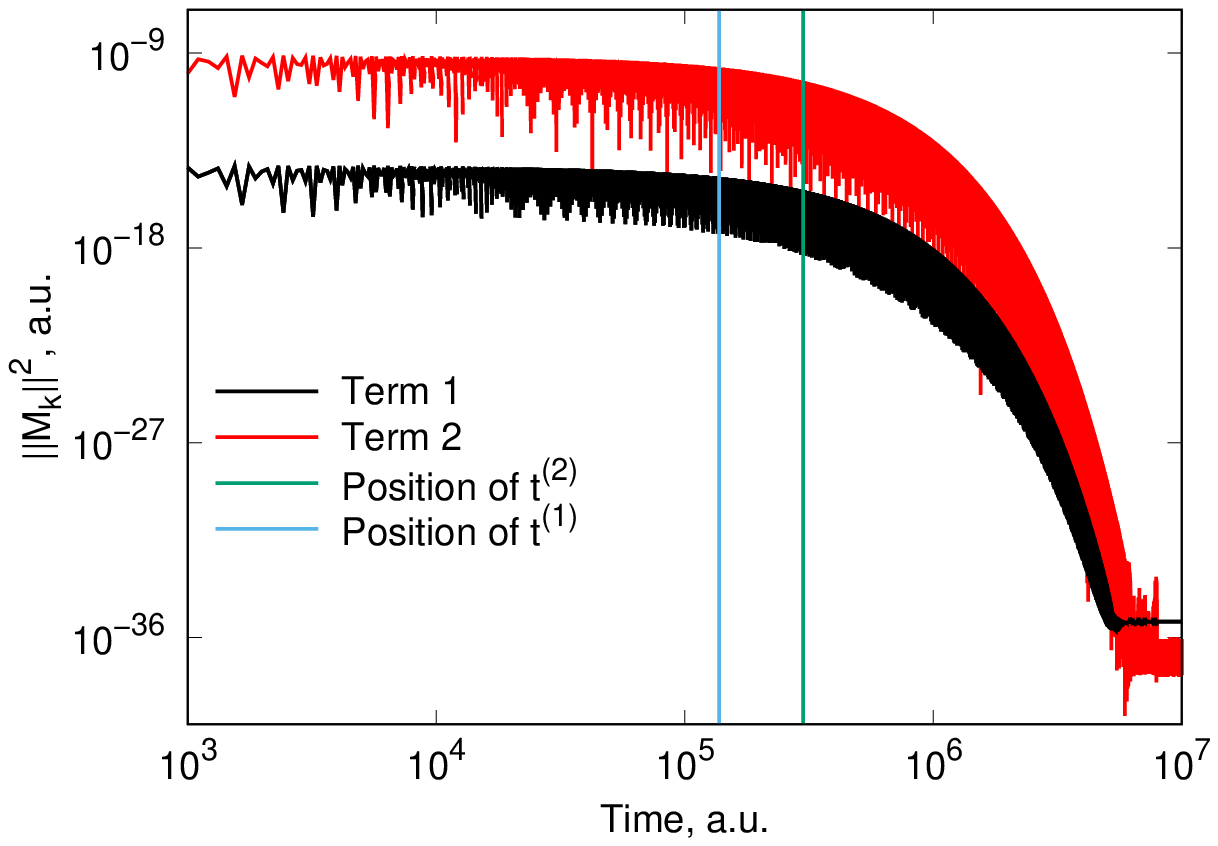} \\
b)\vspace{-0.5cm}&\\&\hspace{-0.5cm}\includegraphics[width=0.5\textwidth]{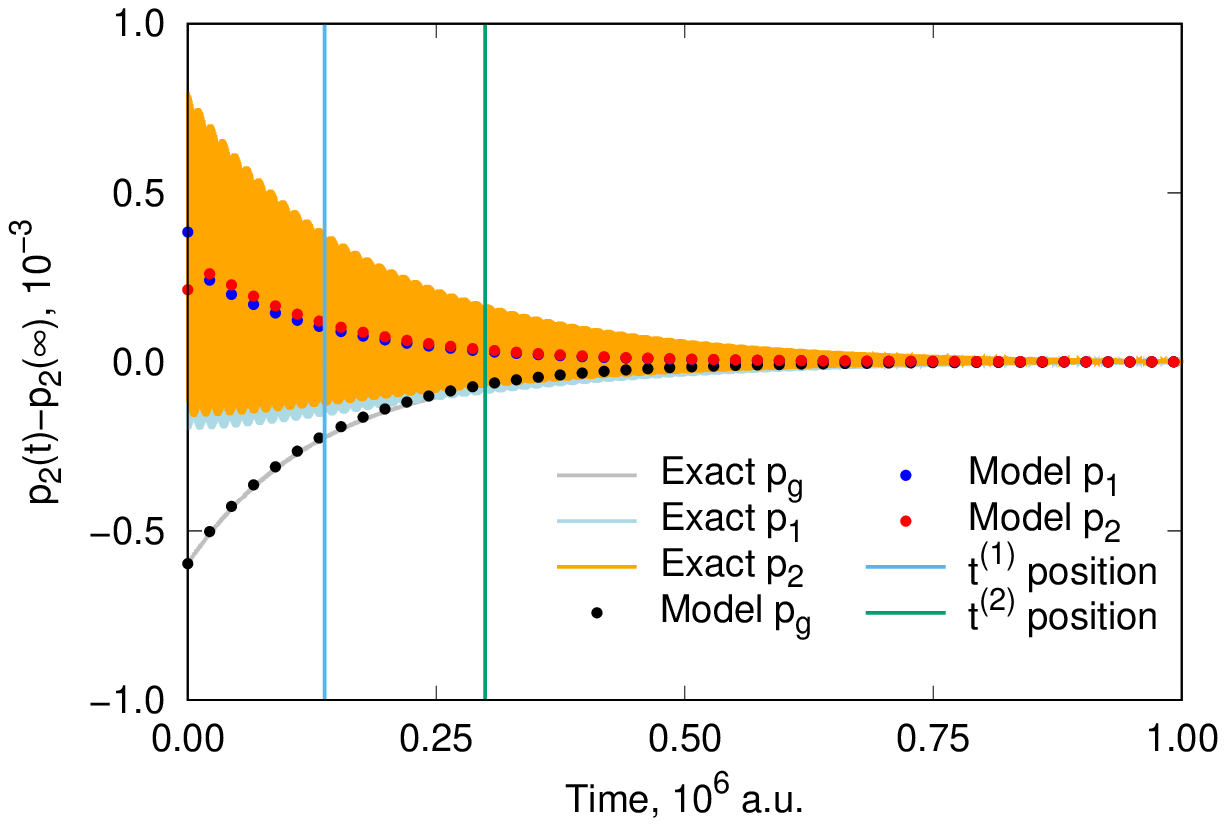}
\end{tabular}
\caption{Time evolution of the V-system in the non-Markovian case. (a) $\mathbf{M}^{(1)}$ and $\mathbf{M}^{(2)}$ and (b) the populations comparing exact propagation and steady state rates 
.}
\protect\label{fig:Dyn_nonMarkov}
\end{figure}


It is clear then that there are circumstances under which perturbations away from the NESS can be used to
obtain information on rates within the NESS. 
These results motivate further, ongoing, work to examine the possibility of identifying physical conditions
under which the Markovian or non-Markovian dynamics applies as a system approaches the NESS, and hence useful conditions
for the utility of perturbations away from the NESS to determine internal rates.

\section{Conclusion and Future work}
\label{sec:Con}

We have presented a rigorous methodology for defining components of a network via projection operators,
and for obtaining rates of population flow between these components in the all important non-equilibrium steady state (NESS).
Quantum effects are included both via
the projection operators as well as in the incorporation of coherences. The versatile projection operator formalism 
allows, for example, the removal of the rate of weak radiative absorption, typically the rate determining step,
exposing the rates within the excited states. 

The {NESS} rates defined in this formalism are not based on the commonly used linear response formalism\cite{Yamamoto1960} and are valid
in all parameter regimes.
In addition, our approach has allowed insights into the temporal range of utility of the kinetic network away from the NESS, in terms of Markovian and
non-Markovian contributions to the time evolution of the populations. This constitutes significant input into the issue of when the NESS rates can
be determined by perturbing the system away from the NESS.
Finally, note that while the examples presented  make use of the Lindblad description of open systems, the methodology is
completely general and can be straightforwardly applied to any description of the Liouville equation, such as
Redfield theory\cite{REDFIELD1965,Egorova2003} or, ideally, the  exact Zwanzig-Nakajima form.

The formalism developed has allowed us to analyze two commonly used population transfer models, the V-system and the nonequilibrium spin-boson model,
giving further insight into the behavior of the {NESS} rates. Both cases where the excitation step is included or excluded were examined.
In addition, the application to the
nonequilibrium spin-boson model with no attenuation of the incident light, showed that
the rate obtained via vertical excitation is approximately three orders of magnitude larger than the {NESS} rate.
This finding supports the view (for a review see Ref.~\citenum{Brumer2018}) that ultra-fast experiments that probe systems that operate naturally
under {NESS} conditions are, in fact, preparing and observing system dynamics that are totally different from those that occur naturally.

The {NESS} rates discussed in this work are a major step forward, since they characterize the rate under proper, natural,
 NESS} conditions, where the network definition is general.
Future work will explore these effects in biologically motivated models such as the Hahn-Stock retinal model.\cite{Hahn2000}
Currently the method relies on the inversion of a Liouville superoperator which is computationally memory intensive for even modestly sized systems.
Future work will focus on ways to alleviate this bottleneck, such as the iterative scheme in Ref. \cite{Axelrod2018},
so that the rate calculation can be applied to larger systems.
In addition to overcoming the computational challenges, the formal connections to other aspects of rate theory will be further explored.

\section*{Acknowledgments}

This material is based upon work supported by the U.S. Air Force Office of Scientific Research under award number FA9550-20-1-0354.


\appendix

\section{Rates by solving a linear equation}
\label{app:rates_lineq}

A simple way to obtain the rates defined by \eq{eq:ourrate} is by solving the system of linear Eqs.~(\ref{eq:sys_of_eqnsP}-\ref{eq:sys_of_eqnsQ}) in the steady state limit.
Taking \eq{eq:sys_of_eqnsQ} for $\rho=\rho_s$, we obtain the linear equation to solve for $\mathcal{Q}[\rho_s]$
\bea
0 &=& \sum_{n=1}^{N} \mathcal{Q}\mathcal{L}[\varrho_n] p_n + \mathcal{Q}\mathcal{L}\mathcal{Q}[\rho_s].
\eea
Writing the solution as
\bea
\mathcal{Q}[\rho_s] &=& -\sum_{n=1}^{N} (\mathcal{Q}\mathcal{L}\mathcal{Q})^{-1}\mathcal{Q}\mathcal{L}[\varrho_n] p_n.
\eea
we can substitute it for $\mathcal{Q}[\rho_s]$ in Eq. (\ref{eq:sys_of_eqnsP}), and obtain
\bea
\dot{p}_m &=& \sum^N_{n=1}\tr\{\hat{P}_m\mathcal{L}[\hat{\varrho}_n]\} p_n \nonumber\\&&
- \sum_{n=1}^{N} \tr\{\hat P_m\mathcal{L}(\mathcal{Q}\mathcal{L}\mathcal{Q})^{-1}\mathcal{Q}\mathcal{L}[\varrho_n]\} p_n,
\eea
which is in fact \eq{eq:red_eqnsp_SS} with our rate definition of \eq{eq:ourrate}.

\section{Rates in the  V-system}
\label{app:proof_V-syst_ungrouped}

Starting from \eq{eq:VsysME}, we wish to derive Eqs.~(\ref{eq:EoMsg}-\ref{eq:EoMs2}) in the steady state limit.
We proceed by solving the system of linear equations~(Eqs. \ref{eq:sys_of_eqnsP}-\ref{eq:sys_of_eqnsQ}) for the populations $\{\rho_{nn}\}$ in the steady state limit (similar to Appendix~\ref{app:rates_lineq}).
Taking matrix elements of \eq{eq:VsysME}, we obtain a set of equations for each element of the density matrix:
\begin{widetext}
\begin{align}
\label{eq:rhoggdot}
\tr\{\op P_g\dot{\op\rho}\}=\dot{\rho}_{gg}
&= -2(\Gamma_{H_1}+\Gamma_{H_2}) \rho_{gg} +2\Gamma_{C_1} \rho_{11} +2\Gamma_{C_2} \rho_{22},
\\\label{eq:rho11dot}
\tr\{\op P_1\dot{\op\rho}\}=\dot{\rho}_{11}
&= -iJ(\rho_{21}-\rho_{12}) +2\Gamma_{H_1} \rho_{gg} -2(\Gamma_{C_1} + \Gamma_{D_f} ) \rho_{11} +2\Gamma_{D_b} \rho_{22},
\\\label{eq:rho22dot}
\tr\{\op P_2\dot{\op\rho}\}=\dot{\rho}_{22}
&= -iJ(\rho_{12}-\rho_{21}) +2\Gamma_{H_2} \rho_{gg} -2(\Gamma_{C_2} +\Gamma_{D_b} ) \rho_{22} +2\Gamma_{D_f} \rho_{11},
%
\\
\label{eq:rhog1dot}
\braOket{g}{\mathcal{Q}[\dot{\op\rho}]}{1}=\dot{\rho}_{g1}
&= i(\epsilon_1-\epsilon_g)\rho_{g1} -(\Gamma_{H_1} +\Gamma_{H_2} +\Gamma_{C_1} +\Gamma_{D_f})\rho_{g1} +iJ\rho_{g2},
\\\label{eq:rhog2dot}
\braOket{g}{\mathcal{Q}[\dot{\op\rho}]}{2}=\dot{\rho}_{g2}
&= i(\epsilon_2-\epsilon_g)\rho_{g2} -(\Gamma_{H_1} +\Gamma_{H_2} +\Gamma_{C_2} +\Gamma_{D_b})\rho_{g2} +iJ\rho_{g1},
\\\label{eq:rho12dot}
\braOket{1}{\mathcal{Q}[\dot{\op\rho}]}{2}=\dot{\rho}_{12}
&= (i\Delta -\Gamma_{C_1} -\Gamma_{C_2} -\Gamma_{D_f} -\Gamma_{D_b}) \rho_{12} +iJ(\rho_{11}-\rho_{22}).
\end{align}
\end{widetext}
The diagonal terms of $\mathcal{Q}[\dot{\op\rho}]$ are trivially zero since $\braOket{k}{\mathcal{Q}\mathcal{L}[{\op\rho}]}{k}=0$,
and equations for other elements can be obtained directly since $\hat{\rho}$ is Hermitian: $\rho_{1g}=\rho_{g1}^*$, $\rho_{2g}=\rho_{g2}^*$, $\rho_{21}=\rho_{12}^*$.

Note first that the elements $\rho_{g1}$ and $\rho_{g2}$ are completely decoupled from the other elements.
Second, from \eq{eq:rho12dot} and its complex conjugate we can express $\rho_{12}-\rho_{21}$. We isolate $\rho_{12}$ as
\bea
\rho_{12} &=& \frac{\dot{\rho}_{12} - iJ(\rho_{11}-\rho_{22})}{i\Delta -\Gamma_{C_1} -\Gamma_{C_2} -\Gamma_{D_f} -\Gamma_{D_b}}.
\eea
In the limit of a steady state, we have that $\dot{\rho}_{12}=0$, which simplifies the expression for $\rho_{12}$:
\bea
\rho_{12} &=& \frac{iJ(\rho_{11}-\rho_{22})}{\Gamma_{C_1} +\Gamma_{C_2} +\Gamma_{D_f} +\Gamma_{D_b} -i\Delta}.
\eea
Hence, we have that
\bea
\rho_{12}-\rho_{21}
&=& \frac{2iJ(\rho_{11}-\rho_{22})(\Gamma_{C_1} +\Gamma_{C_2} +\Gamma_{D_f} +\Gamma_{D_b})}{(\Gamma_{C_1} +\Gamma_{C_2} +\Gamma_{D_f} +\Gamma_{D_b})^2 +\Delta^2}. \nonumber\\
\eea
Substituting the last expression in Eqs.~(\ref{eq:rho11dot}-\ref{eq:rho22dot}), we obtain the set of Eqs.~(\ref{eq:EoMsg}-\ref{eq:EoMs2}).

The transfer of population between states 1 and 2 is mediated by: (i) the dephasing terms $\{\Gamma_{D_f},\Gamma_{D_b}\}$ and (ii) by the imaginary part of the coherence $\rho_{12}$.
Interestingly, the time evolution of $\rho_{12}$ does not depend on $\Gamma_H$, see \eq{eq:rho12dot}, since $\braOket{1}{\mathcal{L}_H}{2}=0$.
Hence,  $\rho_{12}$ does not depend on $\Gamma_H$, nor does the rate between state 1 and state 2.

Note that in the grouping process  $\rho_{12}$ is replaced by terms depending on $\rho_{11}$ and $\rho_{22}$ using the time derivative expression of $\rho_{12}$.

\section{Rates in the grouped V-system}
\label{app:proof_V-syst_grouped}

In this section, we demonstrate how to obtain the rates between two groups: group ``A'', which contains the ground state and state 1, and group 2, which contains only state 2 (as in the previous subsection).
Therefore, we define the projectors $\PL_A$ and $\PL_2$ using the following definitions
\bea
\op P_2 &=& \ket{2}\bra{2}, \\
\op P_A &=& \ket{g}\bra{g}+\ket{1}\bra{1}, \\
\op \varrho_2 &=& \op P_2,\\
\op \varrho_A &=& \sum_{kl\in\{g,1\}}\ket{k}\frac{\rho_{kl}^{(s)}}{\rho_{gg}^{(s)}+\rho_{11}^{(s)}}\bra{l}.\\
\eea

To simplify the derivation, we have in the steady state, $\rho_{g1}^{(s)}=\rho_{g2}^{(s)}=0$, which can be shown by solving the system of Eqs.~(\ref{eq:rhog1dot}-\ref{eq:rhog2dot}) and using the fact that $\Gamma_{H_1}+\Gamma_{H_2}+\Gamma_{C_k}+\gamma_d>0$.
Furthermore, we introduce the variable $r=\rho_{gg}^{(s)}/(\rho_{gg}^{(s)}+\rho_{11}^{(s)})$ to obtain
\bea
\op\varrho_A &=& \ket{g}r\bra{g}+\ket{1}(1-r)\bra{1}.
\eea

The system of equations using the projectors on these two groups is given by
\begin{widetext}
\bea
\label{eq:pAdot}
\dot{p}_A
&=&
2(\Gamma_{C_2}+\gamma_{d_b}) p_2
-2 ((1-r)\gamma_{d_f} + r\Gamma_{H_2}) p_A
-iJ(\rho_{21}-\rho_{12})
\nonumber\\&&
+2 (\gamma_{d_f}-\Gamma_{H_2}) [ (1-r)\rho_{gg} -r\rho_{11} ], \\
\label{eq:p2dot}
\dot{p}_2
&=& 2 ((1-r)\gamma_{d_f} + r\Gamma_{H_2}) p_A - 2(\Gamma_{C_2}+\gamma_{d
_b}) p_2
-iJ(\rho_{12}-\rho_{21})\nonumber\\&&
-2 (\gamma_{d_f}-\Gamma_{H_2}) [ (1-r) \rho_{gg} - r\rho_{11} ], \\
\notag
\braOket{g}{\mathcal{Q}[\dot{\op\rho}]}{g}
&=&  riJ(\rho_{21}-\rho_{12})+ 2( (1-r)(\Gamma_{C_1} + r\gamma_{d_f}) \\
&&- r (\Gamma_{H_1} + (1-r)\Gamma_{H_2}) ) p_A
+ 2((1-r) \Gamma_{C_2} - r\gamma_{d_b}) p_2
\nonumber\\&&
- 2( \Gamma_{H_1} + (1-r) \Gamma_{H_2} + \Gamma_{C_1} + r\gamma_{d_f} ) [ (1-r) \rho_{gg} - r\rho_{11} ],\label{eq:Qggdot}
\\
\braOket{g}{\mathcal{Q}[\dot{\op\rho}]}{1} &=& \dot{\rho}_{g1}, \\
\braOket{g}{\mathcal{Q}[\dot{\op\rho}]}{2}&=& \dot{\rho}_{g2}, \\
\braOket{1}{\mathcal{Q}[\dot{\op\rho}]}{2}&=& \dot{\rho}_{12}.
\eea
\end{widetext}
The other diagonal terms do not introduce more information since $\braOket{1}{\mathcal{Q}\mathcal{L}[{\op\rho}]}{1} = -\braOket{g}{\mathcal{Q}\mathcal{L}[{\op\rho}]}{g}$ and $\braOket{2}{\mathcal{Q}\mathcal{L}[{\op\rho}]}{2}=0$, and other terms are deduced using Hermiticity of $\op\rho$.
To obtain an equation of $p_A$ and $p_2$ only, we need to substitute $\rho_{12}-\rho_{21}$ and $(1-r) \rho_{gg} - r\rho_{11}$ in Eqs.~(\ref{eq:pAdot}-\ref{eq:p2dot}).
The first quantity is replaced using the same procedure as in previous subsection.
Regarding the second quantity, we can isolate it by imposing steady state in \eq{eq:Qggdot} $\braOket{g}{\mathcal{Q}\mathcal{L}[{\op\rho}]}{g}=0$,
\begin{widetext}
\bea\notag
(1-r) \rho_{gg} - r\rho_{11} &=& \\
 \frac{
 ( (1-r)(\Gamma_{C_1} + r(\beta+\gamma_{d_f})) - r (\Gamma_{H_1} + (1-r)\Gamma_{H_2}) ) p_A
+ ((1-r) \Gamma_{C_2} - r(\beta+\gamma_{d_b})) p_2
}{ r(\beta+\gamma_{d_f}) + \Gamma_{H_1} + (1-r) \Gamma_{H_2} + \Gamma_{C_1} }.
\eea
\end{widetext}
Substituting this into Eqs.~(\ref{eq:pAdot}-\ref{eq:p2dot}), we obtain the final set of kinetic equations for the two groups in the steady state
\bea
\dot{p}_A
&=&
-k_{2A}  p_A
+ k_{A2} p_2, \\
\dot{p}_2
&=&
k_{2A} p_A
-k_{A2} p_2,
\eea
where the rates are given by Eqs.~(\ref{eq:k2A}-\ref{eq:kA2}).


%

\end{document}